# Cu$^{2+}$ Affects Amyloid-β (1–42) Aggregation by Increasing Peptide-Peptide Binding Forces

Francis Hane[1], Gary Tran[1], Simon J. Attwood[2], Zoya Leonenko[1,2]*

1 Department of Biology, University of Waterloo, Waterloo, Ontario, Canada, 2 Department of Physics and Astronomy, University of Waterloo, Waterloo, Ontario, Canada

**Abstract**

The link between metals, Alzheimer's disease (AD) and its implicated protein, amyloid-β (Aβ), is complex and highly studied. AD is believed to occur as a result of the misfolding and aggregation of Aβ. The dyshomeostasis of metal ions and their propensity to interact with Aβ has also been implicated in AD. In this work, we use single molecule atomic force spectroscopy to measure the rupture force required to dissociate two Aβ (1–42) peptides in the presence of copper ions, Cu$^{2+}$. In addition, we use atomic force microscopy to resolve the aggregation of Aβ formed. Previous research has shown that metal ions decrease the lag time associated with Aβ aggregation. We show that with the addition of copper ions the unbinding force increases notably. This suggests that the reduction of lag time associated with Aβ aggregation occurs on a single molecule level as a result of an increase in binding forces during the very initial interactions between two Aβ peptides. We attribute these results to copper ions acting as a bridge between the two peptide molecules, increasing the stability of the peptide-peptide complex.

**Funding:** The work was supported by Canadian Foundation of Innovation), Ontario Research Fund and Natural Science and Engineering Council of Canada. The funders had no role in study design, data collection and analysis, decision to publish, or preparation of the manuscript.

**Competing Interests:** ZL is a PLOS ONE Editorial Board member. This does not alter the authors' adherence to all the PLOS ONE policies on sharing data and materials.

* E-mail: zleonenk@uwaterloo.ca

## Introduction

Amyloid-β (Aβ) is a 35–43 long amino acid peptide implicated in the neurodegenerative protein misfolding disease known as Alzheimer's disease (AD) [1]. There are about twenty seven protein misfolding diseases identified including Parkinson's, Huntington's, type II diabetes and protein alveolar proteinosis. Each one of the protein misfolding diseases has an associated protein that misfolds into a pathological state. Normally, Aβ exists primarily as a α-helical or random coil structure, but can misfold into a β-sheet structure that is prone to aggregate into toxic amyloid oligomers and insoluble amyloid fibrils [2,3,4,5]. The mechanism for this misfolding has not yet been identified. The initial misfolding of amyloid-β onto itself occurs through the folding of amino acid sequences 16–23 onto 28–35 to form a β-sheet structure [5]. It is now accepted that the oligomers, which may form along a distinct pathway, are more neurotoxic than the relatively inert amyloid fibrils [6,7,8,9]. Despite extensive research, the mechanism of action of Aβ is not clearly understood.

The factors affecting AD are diverse and their interrelatedness remains elusive. Genetic factors [10], metals [5,11], and vascular deficiencies [12] have been found to be associated with AD. Also, the Alzheimer's afflicted brain has been shown to suffer from severe oxidative stress [13,14] and inflammation [15]. In postmortem brains of AD patients, amyloid plaques were laden with trace metals such as copper, zinc, and iron at concentrations up to 400 μM, 1 mM, and 1 mM, respectively [16]. Extensive research has been conducted on the role of metal ions in the formation of reactive oxygen species (ROS), and amyloid-metal complexes that increase amyloid toxicity by ultimately promoting apoptosis [17,18,19,20,21,22].

Aβ aggregation begins with a lag phase at which point the peptide progressively aggregates to form nucleation seeds [23]. The addition of metal ions has been shown to reduce the lag phase associated with Aβ aggregation [24]. Aβ has been shown to bind metal ions, such as copper, zinc, and aluminum, yielding amyloid-metal complexes with varying effects [16,25,26,27]. The binding of Aβ to copper allows the peptide to insert into lipid membranes more readily [28], while aluminum-Aβ complexes have been shown to disrupt lipid membranes [29].

The binding site of copper is believed to lie within the N-terminal portion of the peptide. Specifically, there is a salt bridge formed utilizing metals, such as zinc and copper, predominantly through a His(13)-metal-His(14) conformation as well as bridges with His(6) [26,30,31].

Previous research has shown that copper binds to these His co-ordination sites with greater affinity than zinc [32] and significantly stabilizes Aβ aggregates [31]. The binding of copper causes Aβ to become redox active, which significantly contributes to the oxidative stress prevalent in AD [5,13,14]. The reduction of Cu$^{2+}$-amyloid complexes to Cu$^{+}$-amyloid complexes has been shown to produce hydrogen peroxide [33] that in turn leads to the formation of pro-apoptotic lipid peroxidation products, such as 4-hydroxynonenal, which ultimately induces neuronal cell apoptosis [18,20]. Thus, the binding of copper to Aβ not only increases neurotoxicity, but it has also been demonstrated to have kinetic and thermodynamic implications [34].

Single molecule atomic force spectroscopy (AFS) in combination with atomic force microscopy (AFM), is a powerful approach to study the effect of metals on amyloid aggregation and can shed light on the very initial step of Aβ aggregation as well as follow the progression of this process with time. Single molecule AFS is an AFM-based technique used to extract information from the interaction of two molecules. Typically, a protein is bound to a substrate and another protein to the tip of an AFM cantilever. The tip is brought in close proximity to the surface and the two molecules are allowed to bind. The tip is then retracted and the peptide-peptide bond ruptures. The AFM apparatus quantitatively measures the rupture force, and this force is recorded for statistical analysis [35,36]. The application of single molecule atomic force spectroscopy to study protein misfolding diseases has been reviewed [37]. In this work, we study the effect of copper ions on the peptide-peptide rupture force of the Aβ (1–42) peptide. We show that when copper ions are added to the Aβ force spectroscopy environment, the rupture force increases dramatically, which correlates with a higher rate of aggregation shown by AFM imaging. This is the first single-molecule study which shows that $Cu^{2+}$ increases the force of interaction between two single Aβ peptides; thus, affecting further aggregation.

## Experimental Procedures

We used a widely accepted method of binding proteins through N-terminus to PEG heterobifunctional cross linkers (35, 36), an experimental setup as previously described [38]. Briefly, the experimental procedures are outlined below.

**Tip Surface and Mica Modification.** Veeco MLCT Silicon Nitride AFM cantilevers were cleaned by soaking in ethanol for 15 minutes, washed in ultrapure water and dried in a gentle stream of nitrogen. The cantilever was then placed under UV light for 30 minutes. Mica was freshly cleaved. 3-aminopropyltriethoxy silane (APS) was synthesized as previously described [39]. The structure of APS was confirmed using NMR spectroscopy. The mica and cantilever were then immersed in 167 µM APS for 30 minutes, then rinsed with ultrapure water and dried under a gentle stream of nitrogen. The mica and cantilever were then placed in a 3400 MW Polyethylene Glycol (PEG) solution (167 µM in DMSO) (Laysan Bio, Alabaster GA) for 3 hours, than rinsed with DMSO. The cantilever and mica were washed and stored in HEPES buffer (50 mM HEPES, 150 mM NaCl, pH 7.4). We specifically chose HEPES buffer because of the absence of metal ions.

**Aβ (1–42) Preparation and Surface Binding.** Cys-Aβ (1–42) was purchased from Anaspec (Fremont, CA) and prepared in DMSO at a concentration of 1 mg/mL. The Aβ stock solution was then diluted in HEPES buffer to a final concentration of 20 nM. An equal volume of 200 nM tris(2-carboxyethyl)phosphine (TCEP) was added to the dilute peptide solution to prevent aggregation. The Aβ solution was stored for 15 minutes and then centrifuged at 14000 RPM for 15 minutes to move monomeric forms to the top of the solution to ensure primarily monomeric forms of the peptide were used. The mica and cantilevers were soaked in the dilute Aβ solution for 30 minutes. The Aβ was rinsed with HEPES buffer, and the mica was treated for 10 minutes with β-mercaptoethanol to react with any available maleimide groups so as to prevent false rupture events. Both cantilever and mica were washed three times with HEPES buffer, and stored in HEPES buffer until use.

**Atomic Force Spectroscopy.** A JPK Nanowizard II atomic force microscope was used for all measurements. Cantilever spring constants were measured using Hutter's thermal tune method [40], which requires both the normal sensitivity and the thermal resonance spectra. The sensitivity was obtained from the gradient of the contact portion of a force-displacement plot acquired on a mica surface. The thermal spectrum was obtained using the JPK hardware. The voltage response of the cantilever deflection measured using the photodiode was converted to units of force by multiplying by the normal sensitivity and the spring constant. Mica coated with Aβ as described earlier was placed on the stage in the liquid cell and immersed in HEPES buffer. A series of force curves were taken with an approach and retract velocity of 400 nm/s. A dwell time of 0.5 seconds was set to allow peptide-peptide binding events. For a single experiment approximately 1000 force curves were recorded, out of which approximately 10% of these showed specific unbinding events. Each experiment was repeated four times with a different cantilever and substrate. For each repeat experiment at least 100 force curves were analyzed, a similar binary distribution was observed and representative experiments are presented. Solutions of $Cu^{2+}$ (purchased from Sigma-Aldrich) in HEPES buffer were prepared at a concentration of 20 nM and were added to the liquid cell for applicable experiments.

**Force Curve Analysis.** JPK data analysis software was used to analyze force curves. A worm like chain (WLC) fit was obtained for each force curve and rupture forces were obtained. Rupture force histograms were fitted with a sum of two Gaussian distributions, and minimized using the Levenberg-Marquardt non-linear least squares fitting routine in Matlab. Errors quoted for the most probable rupture force are evaluated as the standard deviation of each distribution, divided by the square root of the effective number of counts for each distribution (estimate of standard error). The effective number of counts was approximated by multiplying the total number of data points by the area fraction of the given Gaussian distribution.

**Amyloid Incubation for AFM Imaging.** Aβ (1–42) (purchased from rPeptide, Bogarta, GA) was pre-treated according to the Fezoui procedure [41] to ensure the monomeric form. The peptide solutions were prepared by adding HEPES buffer and either $Cu^{2+}$ ions or an equal amount of buffer to produce the copper and control samples, respectively. The final concentration of amyloid-β (1–42) was 55 µM, and the final concentration of $Cu^{2+}$ was 5.5 µM, which yielded a 10:1 amyloid-$Cu^{2+}$ molar ratio. The solutions were incubated at room temperature for 1 hour, 6 hours and 24 hours. 50 µL aliquots were placed onto freshly cleaved mica at the respective times for a 5 minute adsorption period. Excess amyloid solution was then washed with milliQ water and dried with a gentle stream of $N_2$ gas.

**AFM Imaging.** The mica slides with adsorbed amyloid were placed in a JPK Nanowizard II atomic force microscope and imaged in air in Intermittent Contact mode using cantilevers purchased from Nanosensors™ (Non-contact/Tapping™ mode - High resonance frequency; non-coated; tip radius <10 nm). All images were taken with a line rate of 0.5 Hz, and the gains were adjusted to yield maximum image quality. 10×10 µm and 5×5 µm images were taken, and subsequently analyzed using JPK Data Processing Software. Each experiment was repeated at least twice and at least 3 images for each sample.

## Results

We used a combination of single molecule atomic force spectroscopy and atomic force microscopy to probe the single molecule interactions of Aβ in the presence of $Cu^{2+}$ ions. Statistical analysis was completed on force curves to determine the most probable rupture force and Gaussian curve width. Figure 1 illustrates a schematic of the force spectroscopy experimental set up. Notice that Aβ has been bound to both the tip and substrate through APS and a PEG linker, via a cys residue at the N-terminus.

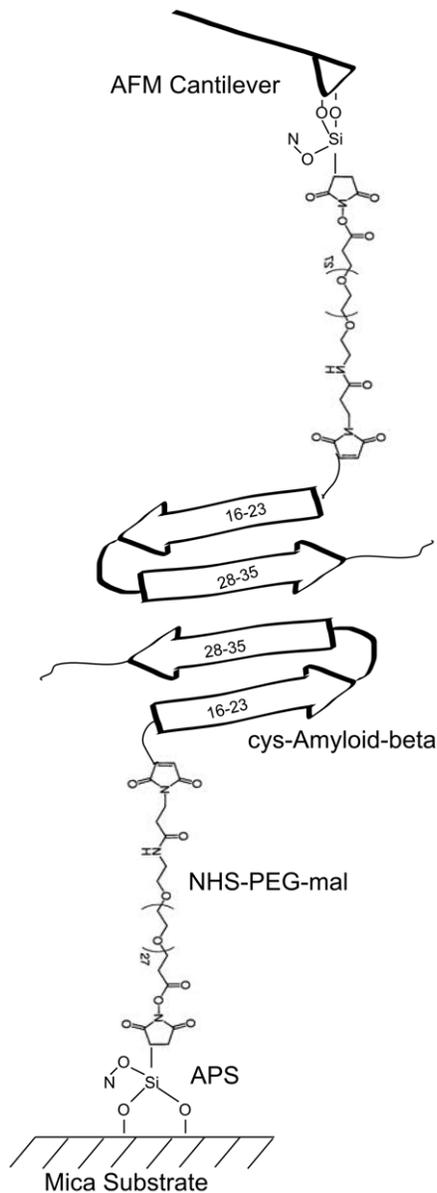

**Figure 1. Force Spectroscopy Setup.** A schematic of experimental setup of force spectroscopy experiment showing Aβ bound to substrate and tip via the PEG linker.
doi:10.1371/journal.pone.0059005.g001

Figure 2 shows a series of histograms of rupture events. For our control experiment without any copper added, we observed double Gaussian peaks centered on 66±1 pN and 132±4 pN. These figures are shown in Table 1. With the addition of copper, a much higher mean rupture force was observed, with copper yielding rupture forces with double Gaussian peaks at 83±3 and 164±5 pN and mean rupture force of 178.9 pN.

Figure 3 shows sample force curves obtained with and without copper and occurring within both the higher and lower Gaussian peaks shown in figure 2.

Figure 4 shows AFM images of amyloid aggregates formed in solution with and without copper ions added for incubation times of 1 hour, 6 hours and 24 hours. We observed oligomeric amyloid species with a mean height of approximately 3.13 nm after an hour of incubation without copper, as seen in Figure 4A. After 6 hours, the control experiment (Figure 4B), revealed the formation of a mixture of oligomeric species, and short fibrils that were approximately 4.5 nm in height. At 24 hours of incubation (Figure 4C), Aβ aggregated to dominantly fibrillar species with a mean height of 7.2 nm. The observed fibrils at 24 hours were significantly longer than those observed at 6 hours, extending up to 3 μm. Figures 4D–4F are representative images of the various structural conformations of the Aβ aggregates in the presence of one tenth $Cu^{2+}$ molar concentrations under the respective times. Large amorphous aggregates with a mean height of 9.3 nm were formed after 6 hours, which coincided with the formation of the short fibrils as found in the respective control. After 24 hours, these unique aggregates remained the dominant species, and the populous fibrils that were observed in the control were not present.

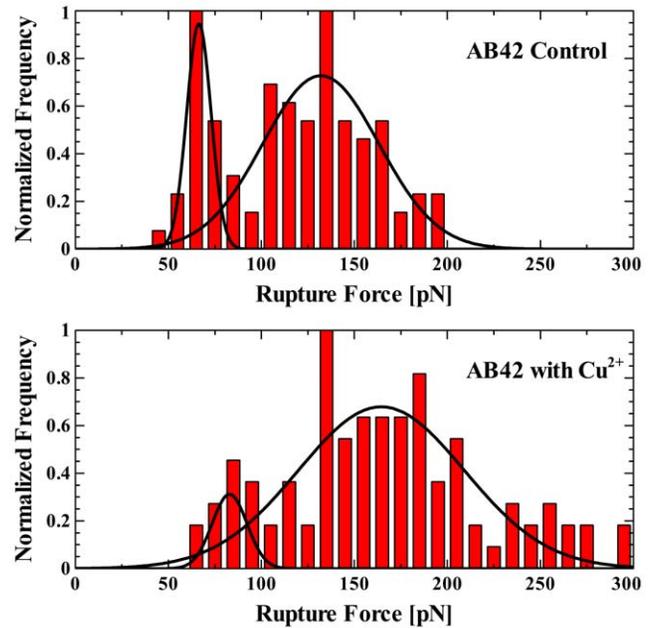

**Figure 2. Effect of Copper on Aβ rupture force.** Histograms show the distribution of forces required to rupture the Aβ-Aβ complex without copper (A) and with copper (B). Fits to the data are Gaussian distributions, the peaks of which represent the most probable rupture force.
doi:10.1371/journal.pone.0059005.g002

## Discussion

In a recent report, Sarell and colleagues [24] attributed the increase in aggregation of Aβ with substoichiometric levels of $Cu^{2+}$ to charge neutralization caused by the binding of copper ions to the copper binding site at the histidine residues resulting in a peptide more prone to self-association. In this work, we present for the first time measurements of the initial single molecule interaction between two Aβ peptides in the presence of $Cu^{2+}$ ions. We show that with the addition of copper, the unbinding (rupture) force increases notably. The increase in unbinding force is consistent with the findings of other groups showing a reduction of lag time in amyloid aggregation when copper is added to the system [34]. This suggests that the reduction of lag time associated with Aβ aggregation occurs on a single molecule level as a result of the very initial dimerization interactions between the peptides.

Table 1. Statistical Data of Force Spectroscopy Experiments.

| | Gaussian Peak 1±SE (pN) | Gaussian Peak 2±SE (pN) | Mean Rupture Force (pN) | Experimental Yield (%) |
|---|---|---|---|---|
| Aβ Control | 66±1 | 132±4 | 125.2 | 14.3 |
| $Cu^{2+}$ added | 83±3 | 164±5 | 178.9 | 14.2 |

doi:10.1371/journal.pone.0059005.t001

The rate of aggregation of aggregation-prone proteins, such as Aβ (1–40) and α-synuclein, has been shown to be a function of the mean rupture force between two peptides [42,43]. It has been established that the acceleration of aggregation is the result of a decrease in the lag time associated with amyloid nucleation [44]. Although the mechanisms involved during this lag time have previously been unclear, it is believed that the lag time is a result of the development of a significant amyloid nucleus onto which other peptides can bind to. We suggest that a reduction in lag time may occur as a result of the very initial nucleation process: the dimerization of two Aβ peptides.

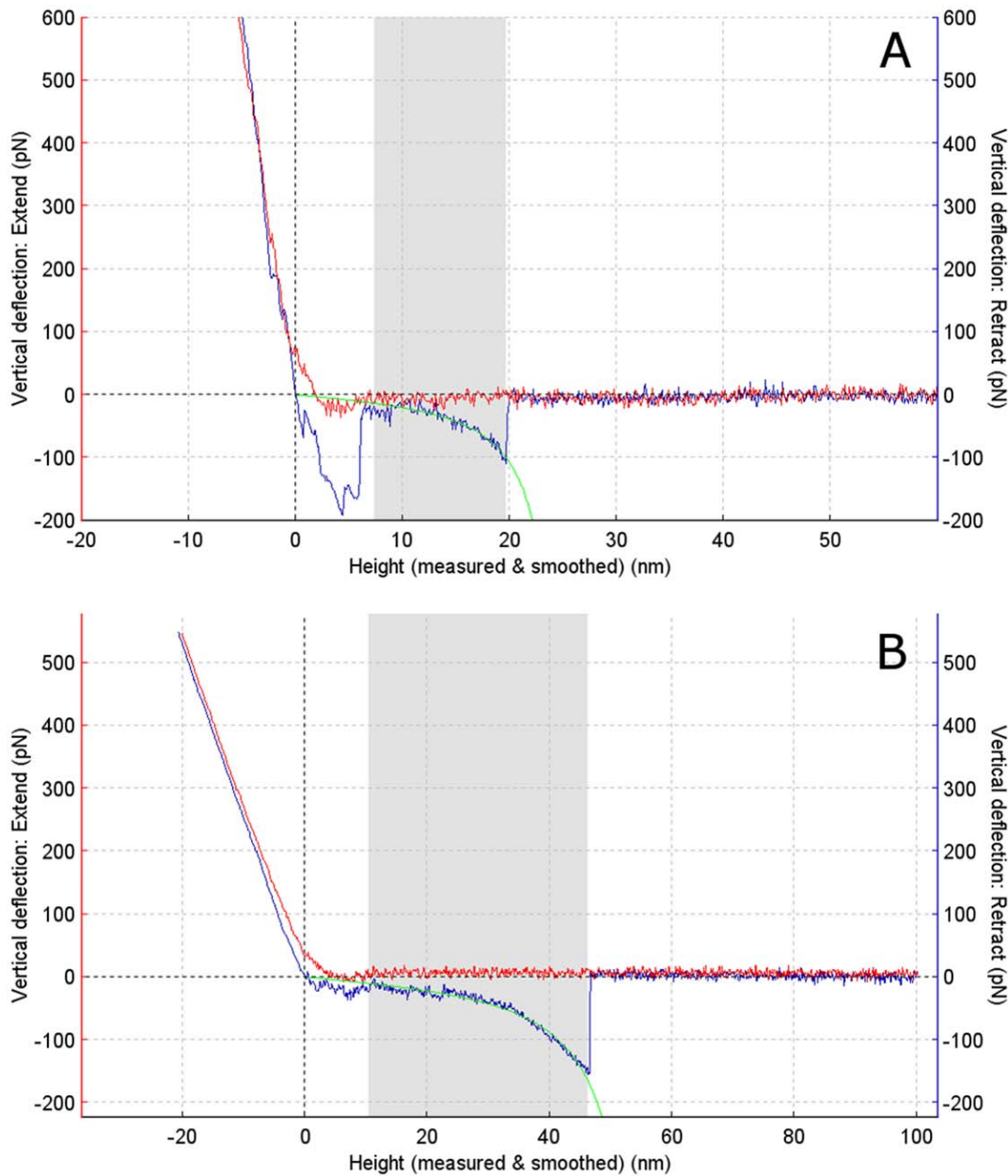

Figure 3. Representative force curves. Force curves showing rupture forces of an Aβ dimer without (A) and with (B) copper added at a retraction rate of 400 nm/s. Curves are shown as force vs. piezo z-displacement.
doi:10.1371/journal.pone.0059005.g003

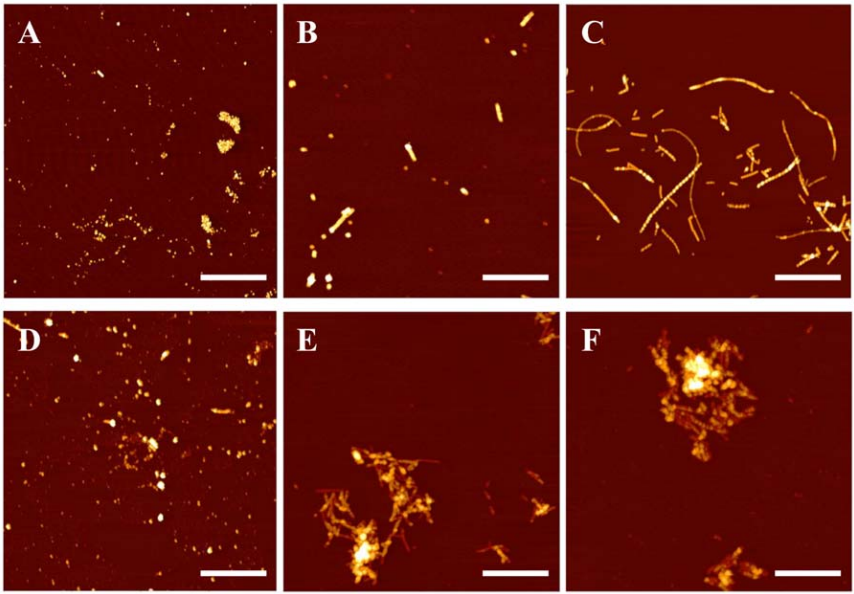

**Figure 4. AFM images of amyloid-metal aggregates.** AFM images of Aβ incubated without copper for periods of 1 hr (A) 6 hr (B) and 24 hr (C), and with copper at a 10:1 molar ratio for 1 hr (D), 6 hr (E), and 24 hr (F). The lateral scale bar is 1 μm.
doi:10.1371/journal.pone.0059005.g004

The common theory of the aggregation of Aβ involves the oligomer cascade hypothesis [1]. According to the oligomer cascade hypothesis, monomeric species form a dimeric nucleation site. Additional monomers are added to this nucleus to progressively form larger oligomers, protofibrils and finally, mature amyloid fibrils [1]. This paradigm has recently come under review and a serious argument can now be made that pathological oligomers and inert fibrils may form along separate pathways [7,38,45]. Necula and colleagues have suggested that several different oligomeric species may form following the misfolding of the Aβ monomer [9]. Only one of these oligomers may eventually form fibrils, with other oligomers remaining in the most stable oligomeric state. He and colleagues [44] studied the amyloid forming β-lactoglobulin protein and have suggested a bifurcation of its amyloid pathway at the 16mer point, where the aggregate may continue as an oligomer or begin to form a protofibril [44]. Based on our data, we propose that, for Aβ, the divergence of this pathway begins much earlier: at the initial dimerization of the two Aβ peptides, where the structure of the initial dimer varies and determines the pathway followed. In our control experiments without copper, we observe two distinct force peaks most likely associated with different dimer configurations, possibly parallel and anti-parallel for two amyloid peptides interacting with each other at the self-recognition site as proposed by Tjernberg [46,47] and illustrated in figures 5A and B. The anti-parallel dimer configuration (Figure 5A) is the more stable of the two stabilized by salt bridges at each end [48], and therefore, we assign this configuration to the stronger force observed (peak two, Figure 2A). The first weaker force (peak one, Figure 2A) more likely corresponds to the parallel configuration (Figure 5B).

Our interpretation of the data collected is consistent with the results of Pedersen and colleagues [49]. Pedersen used bulk measurements of Aβ aggregation under the influence of $Cu^{2+}$ and concluded that copper alters the aggregation pathway of Aβ. The interpretation of our data is consistent with Pedersen's conclusions. Our results build on this data and propose a structural model which is consistent with the observations provided by our groups.

The addition of copper ions significantly increases the unbinding forces of Aβ peptides, at the same time the two distinct peaks shift to a higher value (Figure 2B). It was also apparent that though there is a shift in the two Gaussian peaks, the proportion of the number of binding events in the first peak to the second peak decreases when copper is added to the environment. Based on our hypothesis that there are at least two different conformations of the Aβ dimer, we believe that the addition of copper increases the probability of Aβ to dimerize in a conformation correlating with the second peak, corresponding to a larger binding force. Given these differences in unbinding forces, we suggest that both the parallel and anti-parallel dimer conformations become stabilized by $Cu^{2+}$ ions, which results in the shift of these peaks to higher forces. Considering the possibility of Aβ-Aβ binding both with and without $Cu^{2+}$ ions, we suggest four possible complexes that can be formed in this case: Aβ-Aβ parallel, Aβ-Cu-Aβ parallel, Aβ-Aβ anti-parallel, Aβ-Cu-Aβ antiparallel. Our hypothesis is further supported by recent theoretical work by Mousseau [50] and Urbanc [51]. Using molecular dynamics simulations, both groups independently demonstrated that Aβ (1–42) can dimerize in multiple conformations along multiple pathways. It stands to reason that different dimer conformations have different unbinding forces.

We considered alternative explanations to our hypothesis to explain the presence of two most probable force peaks. We considered that two distinct dimers are rupturing at the same time yielding one much larger force. Based on the probability of a binding and unbinding event, we applied the method proposed by Akhremitchev [52] to determine the probability of two or more peptides being located in the same area and rupturing simultaneously. For our highest yielding experiments, where the most rupture events happened for a given number of approaches, we calculate this probability to be p = 0.12. Consistent with previous analysis methods, the force curves with the (p×n, n = number of force curves) highest rupture forces were discarded and not included in the rupture force histograms. Over several thousand collected force curves, we did indeed observe some double, and even triple, binding events. However, given the inhomogeneity of the length of the PEG linkers, these events were identified by two force curves with the distance between them being a function of the difference in PEG linkers rather than one large force curve. Given our statistical observations, we conclude that any double unbinding event masquerading as a single force curve is so improbable as to be negligible, and certainly would not approach the greater than 50% of force curves that occur at the higher force.

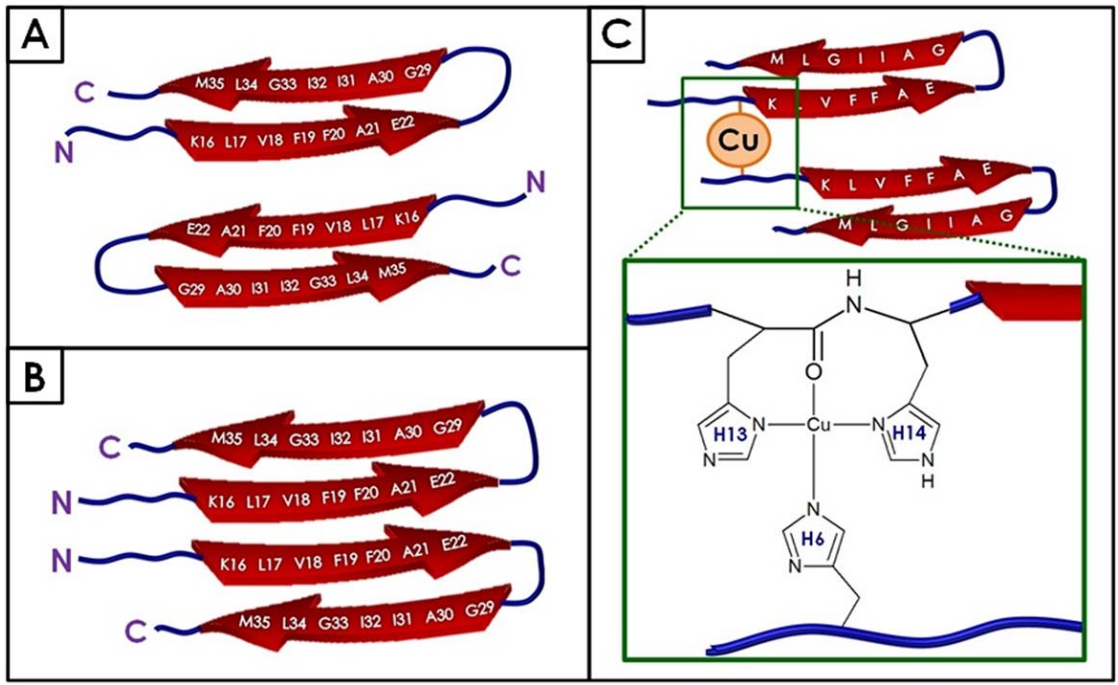

**Figure 5. Schematic diagram of Aβ dimers with and without copper.** Without copper, the most favorable conformation of the Aβ dimer involves an anti-parallel conformation (A). With the addition of copper, Aβ adopts a parallel dimer conformation (B) stabilized by the occupied copper binding sites (C).
doi:10.1371/journal.pone.0059005.g005

Our second alternative explanation for the presence of these double force peaks is that instead of a monomeric peptide being bound at the end of the PEG linker, aggregation has occurred prior to attachment to the PEG linker and an amyloid oligomer was in fact bound at the end of the PEG linker. We closely followed the procedure developed by the Lyubchenko group, who repeatedly showed that when treated as noted in our methods and kept in such a dilute solution, Aβ will not aggregate [38,42,43].

Figure 6 shows plausible conformations of Aβ dimer with and without copper assembled from stable Aβ(1–42) monomer structures (Reference [53] and private discussion with A. Rauk). Each monomer has an internal antiparallel β-sheet between residues 18–21 and 30–33. The dimers are assembled by juxtaposition of the self-recognition site residues 18–21 in antiparallel (A, C) and parallel (B, D) orientation. Both orientations bring His6, His13 and His14 of each monomer into close proximity, requiring little reorientation to bind $Cu^{2+}$ ions (filled green circles). Figure 6 A and B show plausible assemblies for the Aβ-Aβ complexes in antiparallel and parallel conformations without copper. Figure 6 C and D show two possible structures for Aβ-Cu-Aβ in anti-parallel and parallel conformations, respectively. The variety of possible structures and the strength-ening of each binding event by $Cu^{2+}$ results in the broad distribution of unbinding forces we observed in the presence of $Cu^{2+}$ ions (Figure 2B).

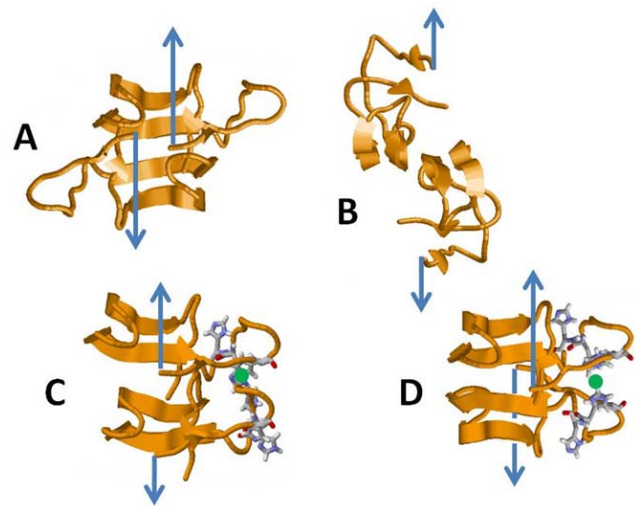

**Figure 6. Proposed structures of Aβ dimers with and without copper assembled from stable Aβ(1–42) monomer structures.** Each monomer has an internal antiparallel β-sheet between residues 18–21 and 30–33. The dimers are assembled by juxtaposition of the self-recognition site residues 18–21 in antiparallel (A, C) and parallel (B, D) orientation. Both orientations bring His6, His13 and His14 of each monomer into close proximity, requiring little reorientation to bind $Cu^{2+}$ ions (filled green circles). Structures are courtesy of D. F. Raffa and A. Rauk, Molecular Dynamics Study of the Beta Amyloid Peptide of Alzheimer's Disease and its Divalent Copper Complexes [53], created using Raswin software.
doi:10.1371/journal.pone.0059005.g006

Our AFM imaging shows that after 6 hours of incubation in the presence of copper, no amyloid fibrils were found in a significant amount. Rather, we observed large amorphous structures that indicate the presence of a copper-dependent Aβ aggregation pathway distinctive from the aggregation pathway without $Cu^{2+}$ which leads to fibril formation. The dominance of these amorphous amyloid aggregates in the presence of copper ions is consistent with previously published images of amyloid aggregation in the presence of $Cu^{2+}$ [54,55].

Trace amounts of metal ions have been shown to decrease the lag time associated with aggregation [34]. Previously, it has been unclear why only substoichiometric amounts of metal ions were needed to reduce the lag time, and thus, increase aggregation. We believe that the reduction of lag time associated with amyloid aggregation in the presence of copper is a result of the very initial dimerization process immediately forming an aggregation nucleus which other peptide can bind onto.

In summary, we demonstrated that unbinding forces of two Aβ peptides without $Cu^{2+}$ have two distinct force peaks, likely associated with parallel and anti-parallel configurations, and resulted in amyloid fibril formations as demonstrated by AFM imaging. The addition of $Cu^{2+}$ ions resulted in a shift to higher force distributions and a higher proportion of unbinding events occurring in the higher force. As seen by the AFM imaging, this distinct force profile is correlated with the formation of amorphous aggregates. We assign this effect of $Cu^{2+}$ to the strengthening of binding between two individual Aβ peptides and disruption of fibril formation pathway at a single molecule level.

## Conclusions

In conclusion, we report that copper increases peptide-peptide binding forces at a single molecule level and changes aggregation observed at the microscale. Therefore, single molecule peptide-peptide interaction defines a pathway for amyloid aggregation. This finding leads to a better understanding of the role of biometals in the mechanism of amyloid fibril formation.

## Acknowledgments

The authors would like to thank Prof. Scott Taylor (University of Waterloo) and his laboratory for synthesis of APS. We also acknowledge Melesa Hane for the critical reading of the manuscript and Brenda Yasie Lee for assistance in creating drawings. Authors greatly appreciate critical discussion of the manuscript and contributed proposed structures of dimers by Prof. Arvi Rauk.

## Author Contributions